\newif\ifDEBUG
\DEBUGtrue

\newif\ifANONYMOUS

\newif\ifARXIV
\ARXIVtrue

\ifARXIV
  \documentclass[sigconf]{acmart}
\else
  \documentclass[sigconf,review]{acmart}
\fi
\ifARXIV
  \setcopyright{none}                    
  \renewcommand\footnotetextcopyrightpermission[1]{}
  \acmConference{}{}{}                   
  \acmBooktitle{}
  \acmYear{}
  \acmISBN{}
  \acmDOI{}
  \acmPrice{}
\else
  \acmConference[JAWs 2026]{Journal Ahead Workshop (JAWs) 2026}{April 2026}{Rio de Janeiro, Brazil}
  \acmBooktitle{Journal Ahead Workshop (JAWs) 2026}
  \acmYear{2026}
  \copyrightyear{2026}
  \acmDOI{}
  \acmISBN{}
  \acmPrice{}
\fi

\usepackage{tabularx}   
\usepackage{booktabs}    
\usepackage{algorithm}
\usepackage[noend]{algpseudocode} 
\usepackage{amsmath}

\usepackage{amssymb}
\usepackage{graphicx}

\usepackage{misc/acmart-taps}
\usepackage[export]{adjustbox} 
\usepackage{amsmath}
\usepackage{array}
\usepackage[USenglish]{babel}
\usepackage{misc/bootstrapicons}
\usepackage{colortbl}
\usepackage{cleveref}
\usepackage{enumerate}
\usepackage{float}
\usepackage{forest}
\usepackage[htt]{hyphenat}
\usepackage{makecell}
\usepackage{multirow}
\usepackage{textgreek}
\usepackage{soul}
\usepackage{pifont}
\usepackage{subcaption} 
\usepackage{url}
\usepackage{xspace}
\usepackage{wrapfig}
\usepackage{paracol}
\usepackage{tabularx}
\usepackage{misc/picins}

\usepackage{tikz} 
\usetikzlibrary{automata} 
\usetikzlibrary{positioning} 
\usetikzlibrary{arrows} 
\usetikzlibrary{calc} 
\usetikzlibrary{patterns} 
\usetikzlibrary{snakes} 
\usetikzlibrary{shapes} 
\usetikzlibrary{arrows.meta} 

\newcommand{\eg}{\textit{e.g.,\ }}

\usepackage{mdframed}
\mdfsetup{skipabove=0.5\topskip,skipbelow=0.5\topskip,align=center}

\usepackage{amsthm}
\newtheorem{thm}{Theorem}

\usepackage{enumitem}
\setlist[itemize]{leftmargin=*,noitemsep,topsep=0pt}
\setlist[enumerate]{leftmargin=*}

\newlist{researchquestions}{enumerate}{1}
\setlist[researchquestions]{label*=\textbf{RQ\arabic*}, leftmargin=*}

\crefformat{section}{\S#2#1#3}
\crefmultiformat{section}{\S#2#1#3}{, \S#2#1#3}{, and \S#2#1#3}{}
\crefname{figure}{Figure}{Figures}
\crefname{appendix}{Appendix}{Appendices}
\crefname{table}{Table}{Tables}
\crefname{algorithm}{Algorithm}{Algorithms}
\crefname{listing}{Listing}{Listings}
\crefname{theorem}{Theorem}{Theorems}
\crefname{thm}{Theorem}{Theorems}
\crefname{lemma}{Lemma}{Lemmata}
\crefname{equation}{Eqt.}{Eqts.}
\crefformat{Grammar}{Grammar #1}

\ifDEBUG
  \PassOptionsToPackage{draft}{commenting}
\else
  \PassOptionsToPackage{final}{commenting}
\fi

\usepackage[nompar]{misc/commenting}

\declareauthor{TS}{Tanmay}{purple}
\declareauthor{JD}{Jamie}{blue}
\declareauthor{BC}{Berk}{orange}
\declareauthor{PA}{Paschal}{magenta}

\authorcommand{TS}{comment}
\authorcommand{JD}{comment}
\authorcommand{BC}{comment}
\authorcommand{PA}{comment}

\onlyauthors{BC,JD,TS,PA}

\usepackage{tcolorbox}

\usepackage{xcolor}

\usepackage{adjustbox}
\usepackage{booktabs}
\usepackage{caption}

\title{Evaluating Dependency Decisions Beyond Functional Correctness in AI Agents}
\title{Towards a Benchmark for Dependency Decision-Making}

\author{Tanmay Singla, Berk \c{C}akar, Paschal C. Amusuo, and James C. Davis}
\orcid{0000-0003-2495-686X}
\affiliation{%
  \institution{Electrical \& Computer Engineering, Purdue University}
  \country{West Lafayette, IN, USA}
}
\email{singlat@purdue.edu, bcakar@purdue.edu, pamusuo@purdue.edu, davisjam@purdue.edu}



\newcommand{\HumanHalNumberTempA}{0}
\newcommand{\HumanHalNumberTempB}{1814}
\newcommand{\HumanHalNumberTempC}{344}
\newcommand{\HumanHalNumberNonTempA}{297}
\newcommand{\HumanHalNumberNonTempB}{465}
\newcommand{\HumanHalNumberNonTempC}{79}
\newcommand{\HumanHalNumberA}{%
  \the\numexpr \HumanHalNumberTempA + \HumanHalNumberNonTempA\relax
}
\newcommand{\HumanHalNumberB}{%
  \the\numexpr \HumanHalNumberTempB + \HumanHalNumberNonTempB\relax
}
\newcommand{\HumanHalNumberC}{%
  \the\numexpr \HumanHalNumberTempC + \HumanHalNumberNonTempC\relax
}

\newcommand{\HumanCritFixCount}{850}
\newcommand{\HumanCritFixPct}{33.3}
\newcommand{\AgentCritFixCount}{229}
\newcommand{\AgentCritFixPct}{15.9}

\newcommand{\HumanHighFixCount}{352}
\newcommand{\HumanHighFixPct}{13.8}
\newcommand{\AgentHighFixCount}{244}
\newcommand{\AgentHighFixPct}{17.0}

\newcommand{\HumanModFixCount}{1,268}
\newcommand{\HumanModFixPct}{49.6}
\newcommand{\AgentModFixCount}{864}
\newcommand{\AgentModFixPct}{60.0}

\newcommand{\HumanLowFixCount}{84}
\newcommand{\HumanLowFixPct}{3.3}
\newcommand{\AgentLowFixCount}{102}
\newcommand{\AgentLowFixPct}{7.1}

\newcommand{\HumanDepAddedUpdated}{40,110}
\newcommand{\AgentDepAddedUpdated}{37,574}
\newcommand{\HumanVulTotal}{659}
\newcommand{\AgentVulTotal}{924}
\newcommand{\HumanVulRateRaw}{1.64\%}
\newcommand{\AgentVulRateRaw}{2.46\%}

\newcommand{\HumanVulRateMitigatable}{83.31\%}
\newcommand{\AgentVulRateMitigatable}{86.58\%}

\newcommand{\HumanIntroRate}{1.03}
\newcommand{\AgentIntroRate}{1.73}

\newcommand{\HumanFixRURate}{5.85}
\newcommand{\AgentFixRURate}{2.82}

\newcommand{\HumanFixRate}{3.10}
\newcommand{\AgentFixRate}{1.55}

\newcommand{\HumanNetImpact}{1,316}
\newcommand{\AgentNetImpact}{98}

\newcommand{\HumanFixTotal}{1975}

\newcommand{\AgentFixTotal}{826}

\newcommand{\HumanMajorFixTotal}{85}
\newcommand{\HumanMajorFixRate}{12.9\%}
\newcommand{\AgentMajorFixTotal}{340}
\newcommand{\AgentMajorFixRate}{36.8\%}

\newcommand{\AgentCritCount}{290}
\newcommand{\AgentCritPct}{12.0}
\newcommand{\HumanCritCount}{305}
\newcommand{\HumanCritPct}{23.3}

\newcommand{\AgentHighCount}{629}
\newcommand{\AgentHighPct}{26.0}
\newcommand{\HumanHighCount}{284}
\newcommand{\HumanHighPct}{21.7}

\newcommand{\AgentModCount}{1,265}
\newcommand{\AgentModPct}{52.3}
\newcommand{\HumanModCount}{588}
\newcommand{\HumanModPct}{44.9}

\newcommand{\AgentLowCount}{234}
\newcommand{\AgentLowPct}{9.7}
\newcommand{\HumanLowCount}{132}
\newcommand{\HumanLowPct}{10.1}

\begin{document}



\begin{abstract}
AI coding agents increasingly modify real software repositories and make dependency decisions, including adding, removing, or updating third-party packages. These choices can materially affect security posture and maintenance burden, yet repository-level evaluations largely emphasize test passing and executability without explicitly scoring whether systems (i) reuse existing dependencies, (ii) avoid unnecessary additions, or (iii) select versions that satisfy security and policy constraints.

We propose \textbf{DepDec-Bench}, a benchmark for evaluating dependency decision-making beyond functional correctness. To ground DepDec-Bench in real-world behavior, we conduct a preliminary study of 117{,}062 dependency changes from agent- and human-authored pull requests across seven ecosystems. We show that coding agents frequently make dependency decisions with security consequences that remain invisible to test-focused evaluation: agents select PR-time known-vulnerable versions (2.46\%) and exhibit net-negative security impact overall (net impact $-98$ vs.\ $+1{,}316$ for humans). These observations inform DepDec-Bench task families and metrics that evaluate safe version selection, reuse discipline, and restraint against dependency bloat alongside test passing.

\end{abstract}

\maketitle

\section{Introduction}

Modern software systems rely heavily on third-party dependencies from public registries such as NPM, PyPI, and Maven Central~\cite{wang2020empirical, latendresse2022not}. 
Dependencies accelerate development, but can also introduce security risks by incorporating vulnerable third-party code~\cite{zimmermann2019small, ohm2020backstabber, decan2018impact}, making dependency management a critical control point for securing modern software systems~\cite{cisa2023oss_sbom_practices,amusuo2025ztd,okafor2022sok}.

As AI coding agents increasingly automate software development tasks~\cite{li2025riseaiteammatessoftware, jimenez2023swe}, they not only generate code, but also make dependency decisions (adding, removing, and updating packages). These choices reflect decisions about non-functional requirements such as security, policy compliance, and long-term maintenance, and can materially change a project’s attack surface and security posture. Yet, existing repository-level benchmarks largely evaluate functional correctness and executability. They do not explicitly assess dependency decision quality, such as whether an agent (i) reuses dependencies already present in the repository, (ii) avoids unnecessary additions, or (iii) selects versions that satisfy PR-time vulnerability and organizational policy constraints~\cite{jimenez2023swe,hai2025impactscontextsrepositorylevelcode,liu2023repobenchbenchmarkingrepositorylevelcode,du2025dependevalbenchmarkingllmsrepository}.

We argue that dependency decision-making should be treated as a first-class capability for repository-level agents, and we propose \textbf{DepDec-Bench}, a benchmark that evaluates this behavior.
Considerations include safe version selection, reuse when available, and restraint against dependency bloat.
DepDec-Bench separates (i) policy-following when constraints are explicit, from (ii) decision quality when constraints are implicit.

While prior work has studied functional failure modes of AI-assisted dependency management~\cite{EndorLabs2025DependencyManagement, Schulman2025Slopsquatting,Schreiber_2025}, it remains unclear how often agents modify dependencies in practice and what security consequences follow. 
To ground DepDec-Bench in real-world behavior, we conduct a preliminary study of dependency changes in agent-authored pull requests. 
Across 2,807 popular GitHub repositories spanning seven ecosystems, we analyze 117,062 dependency changes from 33,596 agent-authored PRs and 6,618 human-authored PRs, labeling vulnerable selections using advisories available at PR-time. 
We find that agents perform version updates more often than humans (25.5\% vs.\ 15.8\%) and, conditional on introducing or updating a dependency, select PR-time known-vulnerable versions more frequently (\AgentVulRateRaw{} vs.\ \HumanVulRateRaw{}). 

These results show that coding agents routinely make dependency decisions with security and maintenance consequences that remain invisible to test-focused evaluation, motivating evaluation targets such as PR-time vulnerability compliance, reuse discipline, and restraint against unnecessary additions.
We use these observations to inform the design of \textbf{DepDec-Bench}, a benchmark for evaluating dependency decision-making, and conclude with a roadmap for its construction and evaluation.

\section{Dependency Decision-making}
\label{sec:Background}

This section motivates dependency decision-making as a core software engineering practice and explains why reliable benchmarks are needed to assess how AI coding agents perform it.

\subsection{Dependency Decision-making in Software Engineering}
Modern software projects are built using third-party packages, incorporated as direct and transitive dependencies~\cite{wittern2016look,synopsys2024ossra}.
These dependencies accelerate development by providing reusable functionality that engineers leverage to implement features and repair defects~\cite{zeng2024surveythirdpartylibrarysecurity, amusuo2025ztd, singla2023llm}.
Direct dependencies are explicitly selected by maintainers, declared in manifest files (\eg \texttt{package.json} for NPM, \texttt{requirements.txt} for Python, \texttt{pom.xml} for Maven)~\cite{npmPackageJson, pipRequirementsFormat}, and fetched from public or private registries.

Completing an engineering task therefore often requires making concrete dependency decisions.
Engineers must decide whether to use an external library at all, which package to select, which version to adopt, whether an existing dependency can be reused instead of introducing a new one, and how to encode the choice in the manifest~\cite{vargas2020dep, anugerah2025surprise}.
In practice, these choices may also reflect organizational constraints, such as selecting packages only from trusted publishers or preferring versions with lower known risk.
Prior work has also shown these decisions can be non-trivial, with engineers having to trade off compatibility against convenience, stability against security~\cite{jafari2023dependencyupdatestrategiespackage,rombaut2024leveragingcrowddependencymanagement}.

Dependency decisions matter: they determine which third-party code, along with its full transitive dependency chain, becomes part of the system.
A dependency can introduce functional bugs, performance regressions, or security vulnerabilities, and it can impose ongoing maintenance costs as maintainers track updates, compatibility changes, and advisories over time~\cite{wang2020empirical, reyes2024breaking}.
Prior work has also shown such decisions are error-prone, especially when delegated to AI coding agents~\cite{EndorLabs2025DependencyManagement, Schulman2025Slopsquatting,Schreiber_2025}.
For these reasons, dependency decision-making is non-trivial and should be made thoughtfully.

\subsection{Limitations of Existing Agent Benchmarks}
AI coding agents are increasingly adopted to complete software engineering tasks~\cite{li2025riseaiteammatessoftware, jimenez2023swe}.
However, many existing repository-level evaluations of these agents center on functional outcomes~\cite{jimenez2023swe,liu2023repobenchbenchmarkingrepositorylevelcode,du2025dependevalbenchmarkingllmsrepository}: whether an agent understands the codebase and produces a solution that completes the task, measured by signals such as compilation success or passing tests.
These benchmarks have been instrumental for measuring code synthesis and repair capabilities, but they generally treat dependency choices, expressed in manifest files, and other non-functional decisions, as incidental artifacts of producing a working solution rather than as first-class decisions to be assessed.

This creates a mismatch between what is evaluated and what matters in production settings.
A dependency change can preserve test passing while still being objectively undesirable: an unnecessary new library may duplicate existing functionality, a version selection may be inconsistent with organizational constraints, or an update may increase future remediation burden.
Because these outcomes are typically orthogonal to immediate correctness, they remain invisible to standard scoring.

\subsection{Challenges of Creating Reliable Decision Benchmarks}
Given its importance, dependency decision-making should be evaluated, alongside functional correctness, to reflect the security and maintenance consequences of agent-driven repository modifications.

However, benchmarking dependency decision-making is challenging for two reasons.
First, it is difficult to \emph{construct} realistic decision tasks: while issues and patches are routinely archived, the underlying tradeoffs and alternatives considered when choosing dependencies are rarely recorded in a structured form.
Second, it is difficult to \emph{score} decisions: unlike task benchmarks with clear oracles (compilation or tests), dependency choices rarely have a single correct answer and must be evaluated relative to context, available alternatives, and the compromises made.
As a result, reliable benchmarks for dependency decision-making remain limited, despite the growing role of agents in making these choices.

\section{Preliminary Study Motivating DepDec-Bench}
\label{sec:Prelim}

To motivate the need for a benchmark on dependency decision-making, we ask two questions:

\begin{enumerate}
\item How often do AI agents make dependency decisions?
\item Do these decisions have practical implications?
\end{enumerate}

\noindent{The answers also inform the benchmark’s design.}

\subsection{Method}

\subsubsection{Experimental Dataset}
We use the AIDev-pop dataset, comprising 33{,}596 agent-authored and 6{,}618 human-authored PRs from 2{,}807 GitHub repositories.
This dataset enables us to study the specific dependency choices AI agents made  when completing the specific tasks and the practical implications of these choices.
As we do not know whether human- and agent-authored PRs address the same tasks or constraints, we provide human-authored PRs as a contextual baseline and acknowledge this dataset’s limitations.

\subsubsection{RQ1: How often do AI agents make dependency decisions?}

We quantify the dependency decisions made by analyzing changes to the dependency manifest file in each pull request.
We consider each change in the dependency manifest file as an explicit decision point where the PR author altered the project’s dependency graph.

For each pull request, we analyze the code changes in the manifest file, identify the associated dependencies and version using regexes specific to the manifest file format, and classify each change as either (i) dependency addition, (ii) dependency removal, or (iii) dependency version update.
We then report (i) the proportion of pull requests that contained dependency changes, and (ii) the distribution of the different change types (additions/removals/updates).

\subsubsection{RQ2: Do these decisions have practical implications?}

For this preliminary study, we measure the security impact of dependency decisions and use the result to approximate practical implications.
To assess security impact, we evaluate if the dependency decision introduced known security vulnerabilities to the application at the time the pull request was submitted.

For each dependency change where a dependency was added or updated, we extract the corresponding package name and version introduced by the pull request and query the ecosyste.ms vulnerability database to identify vulnerabilities that affected the package and were disclosed before the pull request was submitted.
When available, we record the vulnerability's severity and the first available patched version.

Using this information, we compute: (i) the proportion of dependency additions and updates that introduced known vulnerabilities, (ii) the severity distribution of vulnerabilities, (iii) the proportion of vulnerable dependency changes where there was a patched version available at PR-time, (iv) the semantic version difference between the vulnerable and patched version, and (v) the net impact, defined as the difference between the number of dependency changes that introduced vulnerabilities and those that fixed them.

\subsection{Results}

\subsubsection{RQ1: How often do AI agents make dependency decisions?}

\begin{figure}[t]
  \centering
  \includegraphics[width=\linewidth]{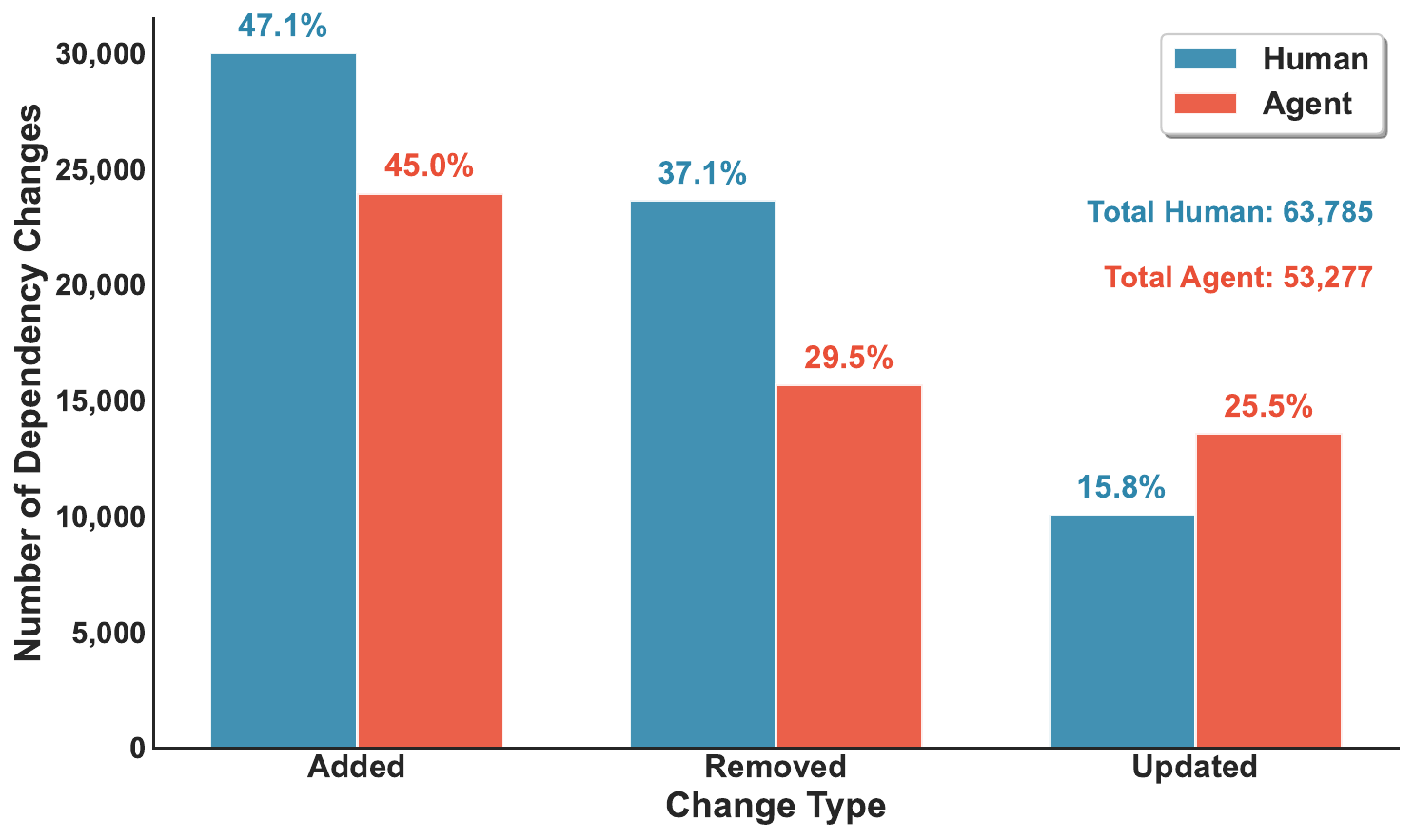}
  \caption{Dependency changes by agents and humans.}
  \label{fig:maintenance-breakdown}
\end{figure}

We identified 117{,}062 dependency changes in the AIDev-pop dataset, of which 53{,}277 occur in 33{,}596 agent-authored pull requests. 
These changes represent observable dependency decisions made by pull request authors while completing tasks.

Classifying each change by type, 45\% are \emph{dependency additions}, 29.5\% are \emph{removals}, and 25.5\% are \emph{version updates} (\cref{fig:maintenance-breakdown}). 
This confirms that coding agents routinely make multiple kinds of dependency decisions: whether to add a new dependency, remove an existing one, and update a dependency’s version.

\textbf{Implication for DepDec-Bench.}
The prevalence of \emph{dependency additions} shows that coding agents frequently face decisions that result in introducing new dependencies. 
However, AIDev-pop does not indicate whether an added dependency was necessary (\eg whether comparable functionality already existed in the repository or could be implemented without a new dependency). 
Our proposed DepDec-Bench therefore includes metrics that measure reuse discipline and reward restraint against avoidable dependency growth, in addition to safe version selection.

\begin{table}[t]
  \centering
\caption{PR-time vulnerability and remediation metrics for dependencies introduced via additions and updates.}
  \label{tab:vuln-metrics}
  \begin{tabular}{lrr}
    \toprule
    \textbf{Metric} & \textbf{Agent} & \textbf{Human} \\
    \midrule
    Dependencies Introduced & \AgentDepAddedUpdated{} & \HumanDepAddedUpdated{} \\
    Vulnerable dependencies & \AgentVulTotal{} (2.5\%) & \HumanVulTotal{} (1.6\%) \\
    Mitigatable (safe version available) (\%) & \AgentVulRateMitigatable{} & \HumanVulRateMitigatable{}  \\
    \midrule
    \multicolumn{3}{l}{\textit{Remediation effort for vulnerable selections}} \\
    Bug-fix (1.2.X $\rightarrow$ 1.2.Y) & 188 (20.3\%) & 266 (40.4\%) \\
    Minor (1.X $\rightarrow$ 1.Y) & 351 (38.0\%) & 283 (43.0\%) \\
    Major (1 $\rightarrow$ 2) & 340 (36.8\%) & 85 (12.9\%) \\
    Other & 45 (4.9\%) & 25 (3.8\%) \\

    \bottomrule
  \end{tabular}
\end{table}

\begin{table*}[t]
  \centering
  \caption{
  Summary of security outcomes by author type.
  Severity labels follow the typical CVSS-to-severity rubric~\cite{ecosystemsAdvisories,cvssSpec,osv,ghsa}.
  }
  \vspace{-0.40cm}
  \label{tab:security-summary}
  \small
  \setlength{\tabcolsep}{5pt}
  \renewcommand{\arraystretch}{0.95}

  \begin{minipage}[t]{0.31\textwidth}
    \centering
    \captionsetup{type=table}
    \caption*{\textbf{(a) Introduced severity}}
    \vspace{-0.25cm}
    \begin{tabular}{lrr}
      \toprule
      \textbf{Severity} & \textbf{Agent} & \textbf{Human} \\
      \midrule
      Critical & \AgentCritCount{} (\AgentCritPct{}\%) & \HumanCritCount{} (\HumanCritPct{}\%) \\
      High     & \AgentHighCount{} (\AgentHighPct{}\%) & \HumanHighCount{} (\HumanHighPct{}\%) \\
      Moderate & \AgentModCount{} (\AgentModPct{}\%)   & \HumanModCount{} (\HumanModPct{}\%) \\
      Low      & \AgentLowCount{} (\AgentLowPct{}\%)   & \HumanLowCount{} (\HumanLowPct{}\%) \\
      \bottomrule
    \end{tabular}
  \end{minipage}
  \hspace{0.0\textwidth}
  \begin{minipage}[t]{0.31\textwidth}
    \centering
    \captionsetup{type=table}
    \caption*{\textbf{(b) Introductions vs.\ fixes}}
    \vspace{-0.25cm}
    \begin{tabular}{lrr}
      \toprule
      \textbf{Category} & \textbf{Agent} & \textbf{Human} \\
      \midrule
      Introduced (\% all changes)       & \AgentIntroRate\%  & \HumanIntroRate\%  \\
      Fixed (\% all changes)            & \AgentFixRate\%    & \HumanFixRate\%    \\
      Fix rate (remove/update)          & \AgentFixRURate\%  & \HumanFixRURate\%  \\
      Net impact (Fixed -- Introduced)  & -\AgentNetImpact{} & +\HumanNetImpact{} \\
      \bottomrule
    \end{tabular}
  \end{minipage}
  \hspace{0.05\textwidth}
  \begin{minipage}[t]{0.31\textwidth}
    \centering
    \captionsetup{type=table}
    \caption*{\textbf{(c) Fixed severity}}
    \vspace{-0.25cm}
    \begin{tabular}{lrr}
      \toprule
      \textbf{Severity} & \textbf{Agent} & \textbf{Human} \\
      \midrule
      Critical  & \AgentCritFixCount{} (\AgentCritFixPct{}\%) & \HumanCritFixCount{} (\HumanCritFixPct{}\%) \\
      High      & \AgentHighFixCount{} (\AgentHighFixPct{}\%) & \HumanHighFixCount{} (\HumanHighFixPct{}\%) \\
      Moderate  & \AgentModFixCount{}  (\AgentModFixPct{}\%)  & \HumanModFixCount{}  (\HumanModFixPct{}\%) \\
      Low       & \AgentLowFixCount{}  (\AgentLowFixPct{}\%)  & \HumanLowFixCount{}  (\HumanLowFixPct{}\%) \\
      \bottomrule
    \end{tabular}
  \end{minipage}
\end{table*}

\subsubsection{RQ2: Do these decisions have practical implications?}

Table~\ref{tab:vuln-metrics} shows that agent dependency decisions introduce PR-time known vulnerabilities more often than human decisions (\AgentVulRateRaw{} vs.\ \HumanVulRateRaw{}). In most of these cases, safer releases existed when the PR was opened (86.58\% had patched versions available) but were not selected, underscoring the relevance of dependency decision-making.

The severity results in Table~\ref{tab:security-summary}(a) indicate that agent-introduced vulnerabilities are concentrated in Moderate and High severities (\AgentModPct{}\% and \AgentHighPct{}\%), while human decisions contain a larger share of Critical cases (\HumanCritPct{}\%). Remediation is also more disruptive for agent selections: Table~\ref{tab:vuln-metrics} shows that \AgentMajorFixRate{} (\AgentMajorFixTotal{}) of agent-introduced vulnerable versions require a major-version upgrade to reach a patched release, compared to \HumanMajorFixRate{} (\HumanMajorFixTotal{}) for human selections, which more often admit non-breaking fixes.

Finally, aggregating outcomes across all dependency edits (Table~\ref{tab:security-summary}(b)), agent-authored PRs fix \AgentFixTotal{} vulnerabilities but introduce \AgentVulTotal{}, yielding a net impact of --\AgentNetImpact{}. Human-authored PRs fix \HumanFixTotal{} and introduce \HumanVulTotal{}, yielding a net impact of +\HumanNetImpact{}. Human fixes also target Critical vulnerabilities more frequently (\HumanCritFixPct{}\% vs.\ \AgentCritFixPct{}\%).

\textbf{Implication for DepDec-Bench.}
Dependency decisions have concrete security consequences: agents often select vulnerable versions despite available patched alternatives, and fixing these choices can require disruptive upgrades. DepDec-Bench should therefore evaluate agents not only on task completion, but also on PR-time vulnerability avoidance and remediation burden.

\section{DepDec-Bench: Design and Roadmap}
\label{DepDec}
Section 2 established that dependency decision-making is a core engineering practice with security and maintenance consequences that are not captured by existing repository-level evaluations. Section 3 showed that coding agents already make such decisions frequently in practice and that these choices can introduce security risk that remains invisible to test-focused scoring. DepDec-Bench is designed to make these decisions observable and measurable.
Since dependency selection has been a longstanding matter of engineering debate, we acknowledge that some of the tasks in the envisioned benchmark may not have specific solutions.
We will have a mix of tasks along a range of subjectivity, supporting the evaluation (or characterization) of software agents as they evolve towards greater autonomy.

DepDec-Bench evaluates dependency decisions through executable repository-level tasks. Each instance uses a pinned repository snapshot and evaluates not only whether tests pass, but also whether the resulting dependency choices are safe, compliant with policy, and disciplined in terms of reuse versus addition. 
We evaluate both
  (1) a \textbf{policy-specified track} (policy provided in the prompt)
  and
  (2) a \textbf{policy-unspecified track} (policy omitted and scored against benchmark-defined rules).
The two tracks distinguish between an agent’s ability to follow explicit dependency constraints and its ability to make sound dependency decisions without such constraints.

DepDec-Bench measures:
\begin{itemize}
    \item \textbf{PR-time safety and policy compliance:} version choices should avoid PR-time known-vulnerable or denylisted releases when compliant alternatives exist.
    \item \textbf{Decision discipline:} agents should reuse appropriate existing dependencies when available and avoid unnecessary additions.
    \item \textbf{Remediation disruption:} when unsafe or non-compliant versions are selected, the induced remediation effort is quantified.
\end{itemize}
These signals define the tasks and metrics used in DepDec-Bench.

\subsection{Task families}
Each DepDec-Bench instance is a repository-level patch synthesis task. The system receives a pinned repository snapshot and a task prompt (optionally with failing-test output) and produces a unified diff patch that may modify source code and dependency manifests. Each instance can be evaluated in both tracks: the same task applies, with or without a specified dependency policy. 

Instances are grouped into four task families:
\begin{enumerate}
    \item \textbf{Reuse-available:} an appropriate dependency is already present; the solution should use it rather than add a new package or reimplement functionality.
    \item \textbf{Justified-add:} the task requires a capability not present in the repository; adding a dependency is appropriate but must comply with an allowlist and version policy.
    \item \textbf{Avoid-unnecessary:} adding a dependency is possible but unjustified; the solution should rely on the standard library or existing project code.
    \item \textbf{Policy-safe selection:} the task requires a version edit; the system must avoid PR-time known-vulnerable or denylisted versions and prefer alternatives available at the reference date.
\end{enumerate}

\noindent\fbox{%
\begin{minipage}{0.97\linewidth}
\textbf{Example tasks}

\begin{itemize}
\item \textbf{Reuse-available (policy-specified).}  
Fix failing ISO-8601 parsing tests.  
\textbf{Policy:} No new dependencies.  
\textbf{Pass:} invokes \texttt{dateutil.parser.isoparse}.  
\textbf{Fail:} adds a parsing dependency or reimplements parsing logic.

\item \textbf{Justified-add (policy-unspecified).}  
Fix webhook signature verification (ECDSA P-256).  
\textbf{Pass:} uses an allowlisted crypto API (\eg \texttt{public\_key.verify(...)} from \texttt{cryptography}).  
\textbf{Fail:} custom crypto/parsing or non-allowed dependencies.  
\textbf{Benchmark scoring:} ad-hoc implementations = unsafe.
\end{itemize}
\end{minipage}}

\subsection{Harness and metrics}
DepDec-Bench uses an executable evaluation harness. Each instance includes (i) a pinned repository snapshot and tests, (ii) a reference date for PR-time advisory lookup, and (iii) an availability definition (\eg present in-tree or in the lockfile). In the policy-specified track, an explicit policy envelope is shown to the system; in the policy-unspecified track, the policy is withheld and defined only by the benchmark. 

The harness applies the patch, installs dependencies under controlled conditions, runs tests, and computes metrics.
Beyond functional success, DepDec-Bench reports:

\begin{itemize}
    \item \textbf{Reuse score:} whether the patch reuses an appropriate existing dependency, credited only when the dependency is observed in execution evidence for the specified test run or entrypoint (\eg import/require traces, dynamic module load logs, etc).
    \item \textbf{Unnecessary-add penalty:} whether the patch introduces new dependencies despite an available solution using existing dependencies or the standard library.
    \item \textbf{Constraint compliance:} whether added or updated dependencies satisfy allowlist and denylist constraints (prompt-provided in policy-specified; benchmark-defined in policy-unspecified).

    \item \textbf{PR-time vulnerability compliance:} whether additions or updates avoid versions known to be vulnerable as of the instance reference date, using an advisory snapshot.
    \item \textbf{Remediation disruption:} if a patch introduces a vulnerable or non-compliant version, the minimal semantic-version change to reach a compliant non-vulnerable release (patch/minor/major).
\end{itemize}

\subsection{Roadmap}
Key next steps include:
\begin{itemize}
    \item \textbf{Build tasks:} (a) constructed tasks with unambiguous dependency decisions, and (b) repository-derived tasks from real changes.
    \item \textbf{Scale construction:} expand instances across ecosystems with standardized manifest parsing and PR-time advisory snapshots, and release an executable harness.
    \item \textbf{Add counterfactual pairs:} create paired instances where only repository context differs (dependency present vs.\ absent) to test reuse and avoidance decisions.
    \item \textbf{Establish baselines:} evaluate frontier models and agentic systems against simple heuristics (prefer-existing, prefer-stdlib, vulnerability avoidance) and run tool-access ablations.
\end{itemize}

\section{Conclusion}

We propose \textbf{DepDec-Bench} to evaluate dependency decision-making in repository-level agents beyond functional correctness, informed by PR-time security signals and practical remediation burden. DepDec-Bench enables reproducible evaluation of dependency decision-making and provides concrete targets for improving agent behavior and CI guardrails for dependency updates.

\textbf{Data Availability:}
Our artifacts (analysis code, processed datasets, and additional visualizations) are available online~\cite{DepDecArtifacts}.

\bibliographystyle{ACM-Reference-Format}
\bibliography{references}

@misc{EndorLabs2025DependencyManagement,
  author       = {{Endor Labs}},
  title        = {State of Dependency Management 2025},
  year         = {2025},
  howpublished = {\url{https://www.endorlabs.com/lp/state-of-dependency-management-2025}},
  note         = {Accessed: 2025-12-01}
}

@online{Schulman2025Slopsquatting,
  author  = {Elad Schulman},
  title   = {Hallucinated Code, Real Threat: How Slopsquatting Targets AI-Assisted Development},
  year    = {2025},
  url     = {https://sdtimes.com/coding-assistants/hallucinated-code-real-threat-how-slopsquatting-targets-ai-assisted-development/},
  note    = {SD Times, Accessed: 2025-12-01}
}

@misc{li2025riseaiteammatessoftware,
      title={The Rise of AI Teammates in Software Engineering (SE) 3.0: How Autonomous Coding Agents Are Reshaping Software Engineering}, 
      author={Hao Li and Haoxiang Zhang and Ahmed E. Hassan},
      year={2025},
      eprint={2507.15003},
      archivePrefix={arXiv},
      primaryClass={cs.SE},
      url={https://arxiv.org/abs/2507.15003}, 
}

@inbook{Schreiber_2025,
   title={Security Vulnerabilities in AI-Generated Code: A Large-Scale Analysis of Public GitHub Repositories},
   ISBN={9789819535378},
   ISSN={1611-3349},
   url={http://dx.doi.org/10.1007/978-981-95-3537-8_9},
   DOI={10.1007/978-981-95-3537-8_9},
   booktitle={Information and Communications Security},
   publisher={Springer Nature Singapore},
   author={Schreiber, Maximilian and Tippe, Pascal},
   year={2025},
   month=oct, pages={153–172} }

@misc{jafari2023dependencyupdatestrategiespackage,
      title={Dependency Update Strategies and Package Characteristics}, 
      author={Abbas Javan Jafari and Diego Elias Costa and Emad Shihab and Rabe Abdalkareem},
      year={2023},
      eprint={2305.15675},
      archivePrefix={arXiv},
      primaryClass={cs.SE},
      url={https://arxiv.org/abs/2305.15675}, 
}

@misc{rombaut2024leveragingcrowddependencymanagement,
      title={Leveraging the Crowd for Dependency Management: An Empirical Study on the Dependabot Compatibility Score}, 
      author={Benjamin Rombaut and Filipe R. Cogo and Ahmed E. Hassan},
      year={2024},
      eprint={2403.09012},
      archivePrefix={arXiv},
      primaryClass={cs.SE},
      url={https://arxiv.org/abs/2403.09012}, 
}

@misc{ecosystemsAdvisories,
  title        = {advisories (ecosyste.ms Data Service)},
  author       = {{Open Source Security Foundation}},
  howpublished = {\url{https://docs.ecosyste.ms/docs/services/data-services/advisories/}},
  note         = {Accessed: 2025-12-21},
  organization = {ecosyste.ms}
}

@techreport{cvssSpec,
  title        = {Common Vulnerability Scoring System (CVSS) Version 4.0: Specification Document},
  institution  = {FIRST.org, Inc.},
  number       = {Document Version 1.2},
  year         = {2024},
  author       = {{FIRST.org}},
  month        = jun,
  howpublished = {\url{https://www.first.org/cvss/v4-0/cvss-v40-specification.pdf}},
  note         = {Accessed: 2025-12-21}
}

@misc{osv,
  title        = {OSV: Open Source Vulnerabilities},
  author       = {{Open Source Security Foundation}},
  year         = {2024},
  howpublished = {\url{https://osv.dev/}},
  note         = {Accessed: 2025-12-21},
  organization = {OSV}
}

@misc{ghsa,
  title        = {About the GitHub Advisory Database},
  howpublished = {\url{https://docs.github.com/code-security/security-advisories/working-with-global-security-advisories-from-the-github-advisory-database/about-the-github-advisory-database}},
  year         = {2025},
  author       = {{GitHub}},
  note         = {Accessed: 2025-12-21},
  organization = {GitHub Docs}
}

@inproceedings{decan2018impact,
  title={On the impact of security vulnerabilities in the npm package dependency network},
  author={Decan, Alexandre and Mens, Tom and Constantinou, Eleni},
  booktitle={Proceedings of the 15th international conference on mining software repositories},
  pages={181--191},
  year={2018}
}

@inproceedings{ohm2020backstabber,
  title={Backstabber’s knife collection: A review of open source software supply chain attacks},
  author={Ohm, Marc and Plate, Henrik and Sykosch, Arnold and Meier, Michael},
  booktitle={International Conference on Detection of Intrusions and Malware, and Vulnerability Assessment},
  pages={23--43},
  year={2020},
  organization={Springer}
}

@inproceedings{zimmermann2019small,
  title={Small world with high risks: A study of security threats in the npm ecosystem},
  author={Zimmermann, Markus and Staicu, Cristian-Alexandru and Tenny, Cam and Pradel, Michael},
  booktitle={28th USENIX Security symposium (USENIX security 19)},
  pages={995--1010},
  year={2019}
}

@inproceedings{wang2020empirical,
  title={An empirical study of usages, updates and risks of third-party libraries in java projects},
  author={Wang, Ying and Chen, Bihuan and Huang, Kaifeng and Shi, Bowen and Xu, Congying and Peng, Xin and Wu, Yijian and Liu, Yang},
  booktitle={2020 IEEE International Conference on Software Maintenance and Evolution (ICSME)},
  pages={35--45},
  year={2020},
  organization={IEEE}
}

@inproceedings{latendresse2022not,
  title={Not all dependencies are equal: An empirical study on production dependencies in npm},
  author={Latendresse, Jasmine and Mujahid, Suhaib and Costa, Diego Elias and Shihab, Emad},
  booktitle={Proceedings of the 37th IEEE/ACM International Conference on Automated Software Engineering},
  pages={1--12},
  year={2022}
}

@article{jimenez2023swe,
  title={Swe-bench: Can language models resolve real-world github issues?},
  author={Jimenez, Carlos E and Yang, John and Wettig, Alexander and Yao, Shunyu and Pei, Kexin and Press, Ofir and Narasimhan, Karthik},
  journal={arXiv preprint arXiv:2310.06770},
  year={2023}
}

@misc{cisa2023oss_sbom_practices,
  author = {{Cybersecurity and Infrastructure Security Agency (CISA)}},
  title  = {Securing the Software Supply Chain: Recommended Practices for Managing Open-Source Software and Software Bill of Materials},
  year   = {2023},
  url    = {https://www.cisa.gov/sites/default/files/2023-12/ESF_SECURING_THE_SOFTWARE_SUPPLY_CHAIN%20RECOMMENDED%20PRACTICES%20FOR%20MANAGING%20OPEN%20SOURCE%20SOFTWARE%20AND%20SOFTWARE%20BILL%20OF%20MATERIALS.pdf},
  note   = {Accessed: 2025-12-23}
}

@misc{npmPackageJson,
  author = {{npm Documentation}},
  title  = {package.json},
  year   = {2025},
  url    = {https://docs.npmjs.com/cli/v10/configuring-npm/package-json},
  note   = {Accessed: 2025-12-23}
}

@misc{pipRequirementsFormat,
  author = {{pip Documentation}},
  title  = {Requirements File Format},
  year   = {2025},
  url    = {https://pip.pypa.io/en/stable/reference/requirements-file-format/},
  note   = {Accessed: 2025-12-23}
}

@inproceedings{okafor2022sok,
  title={Sok: Analysis of software supply chain security by establishing secure design properties},
  author={Okafor, Chinenye and Schorlemmer, Taylor R and Torres-Arias, Santiago and Davis, James C},
  booktitle={Proceedings of the 2022 ACM Workshop on Software Supply Chain Offensive Research and Ecosystem Defenses},
  pages={15--24},
  year={2022}
}

@misc{liu2023repobenchbenchmarkingrepositorylevelcode,
      title={RepoBench: Benchmarking Repository-Level Code Auto-Completion Systems}, 
      author={Tianyang Liu and Canwen Xu and Julian McAuley},
      year={2023},
      eprint={2306.03091},
      archivePrefix={arXiv},
      primaryClass={cs.CL},
      url={https://arxiv.org/abs/2306.03091}, 
}

@misc{hai2025impactscontextsrepositorylevelcode,
      title={On the Impacts of Contexts on Repository-Level Code Generation}, 
      author={Nam Le Hai and Dung Manh Nguyen and Nghi D. Q. Bui},
      year={2025},
      eprint={2406.11927},
      archivePrefix={arXiv},
      primaryClass={cs.SE},
      url={https://arxiv.org/abs/2406.11927}, 
}

@misc{du2025dependevalbenchmarkingllmsrepository,
      title={DependEval: Benchmarking LLMs for Repository Dependency Understanding}, 
      author={Junjia Du and Yadi Liu and Hongcheng Guo and Jiawei Wang and Haojian Huang and Yunyi Ni and Zhoujun Li},
      year={2025},
      eprint={2503.06689},
      archivePrefix={arXiv},
      primaryClass={cs.SE},
      url={https://arxiv.org/abs/2503.06689}, 
}

@misc{DepDecArtifacts,
  title        = {DepDec-Bench Artifacts (code, processed datasets, and visualizations)},
  howpublished = {\url{https://anonymous.4open.science/r/agent_supply_chain_analysis-2FBB/}},
  note         = {Anonymous repository},
  year         = {2026}
}

@misc{zeng2024surveythirdpartylibrarysecurity,
      title={A Survey of Third-Party Library Security Research in Application Software}, 
      author={Jia Zeng and Dan Han and Yaling Zhu and Yangzhong Wang and Fangchen Weng},
      year={2024},
      eprint={2404.17955},
      archivePrefix={arXiv},
      primaryClass={cs.SE},
      url={https://arxiv.org/abs/2404.17955}, 
}

@inproceedings{amusuo2025ztd,
  title={ZTD\_ $\{$JAVA$\}$: Mitigating Software Supply Chain Vulnerabilities via Zero-Trust Dependencies},
  author={Amusuo, Paschal and Robinson, Kyle A and Singla, Tanmay and Peng, Huiyun and Machiry, Aravind and Torres-Arias, Santiago and Simon, Laurent and Davis, James C},
  booktitle={2025 IEEE/ACM 47th International Conference on Software Engineering (ICSE)},
  pages={685--685},
  year={2025},
  organization={IEEE Computer Society}
}

@inproceedings{singla2023llm,
author = {Singla, Tanmay and Anandayuvaraj, Dharun and Kalu, Kelechi G. and Schorlemmer, Taylor R. and Davis, James C.},
title = {An Empirical Study on Using Large Language Models to Analyze Software Supply Chain Security Failures},
year = {2023},
isbn = {9798400702631},
publisher = {Association for Computing Machinery},
address = {New York, NY, USA},
url = {https://doi.org/10.1145/3605770.3625214},
doi = {10.1145/3605770.3625214},
booktitle = {Proceedings of the 2023 Workshop on Software Supply Chain Offensive Research and Ecosystem Defenses},
pages = {5–15},
numpages = {11},
location = {Copenhagen, Denmark},
series = {SCORED '23}
}

@inproceedings{vargas2020dep,
author = {Larios Vargas, Enrique and Aniche, Maur\'{\i}cio and Treude, Christoph and Bruntink, Magiel and Gousios, Georgios},
title = {Selecting third-party libraries: the practitioners’ perspective},
year = {2020},
isbn = {9781450370431},
publisher = {Association for Computing Machinery},
address = {New York, NY, USA},
url = {https://doi.org/10.1145/3368089.3409711},
doi = {10.1145/3368089.3409711},
booktitle = {Proceedings of the 28th ACM Joint Meeting on European Software Engineering Conference and Symposium on the Foundations of Software Engineering},
pages = {245–256},
numpages = {12},
location = {Virtual Event, USA},
series = {ESEC/FSE 2020}
}

@article{anugerah2025surprise,
  title={The Surprise of Multiple Dependency Graphs: Dependency resolution is not deterministic.},
  author={Anugerah, Josie and Martin-Jones, Eve},
  journal={Queue},
  volume={23},
  number={1},
  pages={64--84},
  year={2025},
  publisher={ACM New York, NY, USA}
}

@inproceedings{reyes2024breaking,
  title={Breaking-Good: Explaining Breaking Dependency Updates with Build Analysis},
  author={Reyes, Frank and Baudry, Benoit and Monperrus, Martin},
  booktitle={2024 IEEE International Conference on Source Code Analysis and Manipulation (SCAM)},
  pages={36--46},
  year={2024},
  organization={IEEE}
}

@inproceedings{wittern2016look,
  title={A look at the dynamics of the JavaScript package ecosystem},
  author={Wittern, Erik and Suter, Philippe and Rajagopalan, Shriram},
  booktitle={Proceedings of the 13th international conference on mining software repositories},
  pages={351--361},
  year={2016}
}

@techreport{synopsys2024ossra,
  author      = {{Synopsys Cybersecurity Research Center (CyRC)}},
  title       = {2024 Open Source Security and Risk Analysis (OSSRA) Report},
  institution = {Synopsys, Inc.},
  year        = {2024},
  url         = {https://www.synopsys.com/software-integrity/resources/analyst-reports/open-source-security-risk-analysis.html}
}

\end{document}